\documentclass{aastex}
\usepackage{spr-astr-addons}
\usepackage{url}
\usepackage{epsf}

\RequirePackage{color}

\begin{document}

\title{Dilatonic Equation of Hydrostatic Equilibrium and Neutron Star Structure}
\slugcomment{Not to appear in Nonlearned J., 45.}
%% Running heads
\shorttitle{Dilatonic Equation of Hydrostatic Equilibrium and
Neutron Star Structure} \shortauthors{Hendi et al.}

\author{S. H. Hendi\altaffilmark{1,2}} \and \author{G. H. Bordbar\altaffilmark{1,2}}  \and \author{B. Eslam Panah\altaffilmark{1}}  \and \author{M. Najafi\altaffilmark{1}}
\altaffiltext{1}{Physics Department and Biruni Observatory,
College of Sciences, Shiraz University, Shiraz 71454, Iran}
\altaffiltext{2}{Research Institute for Astronomy and Astrophysics
of Maragha (RIAAM), Maragha, Iran}
%\email{\emaila}

\begin{abstract}
In this paper, we present a new hydrostatic equilibrium equation
related to dilaton gravity. We consider a spherical symmetric
metric to obtain the hydrostatic equilibrium equation of stars in
$4$-dimensions, and generalize TOV equation to the case of
regarding a dilaton field. Then, we calculate the structure
properties of neutron star using our obtained hydrostatic
equilibrium equation employing the modern equations of state of
neutron star matter derived from microscopic calculations. We show
that the maximum mass of neutron star depends on the parameters of
dilaton field and cosmological constant. In other words, by
setting the parameters of new hydrostatic equilibrium equation, we
calculate the maximum mass of neutron star.
\end{abstract}


\section{Introduction}

Our observations of Supernovae type Ia
\cite{Riess,PerlmutterI,PerlmutterII} confirm that the expansion
of our Universe is currently undergoing a period of acceleration.
But Einstein (EN) gravity can not explain this acceleration. In
addition, although Einstein's theory can explain the solar system
phenomena successfully, when we want to study beyond the solar
system or when the gravity is so strong, this theory encounters
with some problems, and so we need to modify EN gravity. In order
to improve EN gravity, one may add a (cosmological) constant to
its Lagrangian \cite{Padmanabhan,Frieman}. Moreover, we can regard
other modifications of Einstein gravity such as Lovelock gravity
\cite{LovelockI,LovelockII,Deruelle,HendiD}, brane world cosmology
\cite{Demetrian,Brax,Gergely,Amarilla}, scalar-tensor theories
\cite{Jordan,Brans,Cai,Fujii,Sotiriou,Giddings,Gregory,Ling,Klepac,Ghodrati}
and $F(R)$ gravity \cite
{Bamba,Cognola,Corda,SotiriouF,Nojiri,HendiEM,HendiES,MomeniRM}.

On the other hand, dark energy and dark matter have received a lot
of attention in recent years. Theoretical physicists introduced
non--baryonic \cite{Dick} dark matters with three models of cold,
warm and hot. Among them cold dark matter model has the highest
agreement with the experimental observations. It is worthwhile to
mention that, dilaton field is one of the most interesting
candidates for cold dark matter \cite{Cho}. In addition, one of
the the best approaches for finding the nature of dark energy is
taking into account a new scalar field \cite{HuangI,HuangII}.
Moreover, the low energy limit of string theory contains a dilaton
field which is coupled to gravity. Physical properties,
thermodynamics, and thermal stability of the black object
solutions in the context of dilaton theory
have been investigated before \cite%
{Tamaki,Yamazaki,Yazadjiev,Dehghani}.

The hydrostatic equilibrium equation (HEE) plays crucial role in studying
the evolution of the stars. This equation is giving an insight regarding the
equilibrium state between internal pressure and gravitational force\ of the
stars.

It is important to note that the neutron and quark stars have
large amount of mass concentrated in small radius. Therefore, they
are in the category of highly dense objects, the so-called compact
stars. Due to this fact, we need to take into account the effects
of general relativity such as the curvature of spacetime for
studying the compact stars. The first HEE in $4$-dimensional
Einstein gravity was studied by Tolman, Oppenheimer and Volkoff
(TOV) \cite{TolmanI,TolmanII,Oppenheimer}. Also, the physical
characteristics of stars using TOV equation have been investigated
in Refs.
\cite{Silbar,Narain,BordbarH,BordbarBY,Li,BordbarR,Yazdizadeh,Oliveira}.
On the other hand, if one is interested in studying the structure
and evolution of stars in different gravities, one should obtain
the HEE in those gravity models. In recent years, the
generalizations and modifications
of this equation were of special interests for many authors \cite%
{Heydarifard,AstashenokCO,Orellana,Arbanil,Doneva,Goswami,Lemos,YazadjievD,CapozzielloDOS,CapozzielloDDFO,MomeniM}.
For more examples, we may note that the HEE equation in $f(R)$ and
$f(G)$ gravities werediscussed in \cite
{AstashenokCOIII,MomeniGMM,AstashenokCOIIII,AbbasMAMQ},
$d$-dimensional HEE in EN gravity was investigated in \cite{Ponce}
and HEE of EN-$\Lambda $ gravity with arbitrary dimensions was
obtained in \cite{BordbarHE}. In addition, $5$ and higher
dimensional HEE in context of Gauss-Bonnet (GB) gravity was
extracted in \cite{Zhan-Ying,Hansraj,BordbarHE}. Recently, $(2+1)
$-dimensional HEE was obtained for a static star in the presence
of cosmological constant \cite{Diaz}.

In this paper, we want to obtain modified HEE in the presence of dilaton
field. We consider the Lagrangian of Einstein-dilaton gravity and a perfect
fluid energy-momentum tensor with spherical symmetric metric to obtain
dilatonic HEE. We also consider dilaton field as a correction of Einstein
gravity to obtain a perturbative HEE. In other words, we obtain HEE for
Einstein gravity with an additional term which the trace of small dilaton
field.

\section{Equation of hydrostatic equilibrium with a dilaton field}

The action of dilaton gravity in the context of Einstein gravity is given by
\begin{equation}
I_{G}=\frac{1}{16\pi }\int d^{4}x\sqrt{-g}\left[R-2 g^{\mu
\nu}\partial_{\mu}\Phi \partial_{\nu}\Phi -V(\Phi) \right]+I_{M},
\label{actEN}
\end{equation}
where $R$ and $\Phi $ are, respectively, the Ricci scalar and the dilaton
field. Also $V(\Phi)$ is the potential for $\Phi $, and $I_{M}$ denotes the
action of matter field which we consider a perfect fluid. Varying the action
(\ref{actEN}) with respect to the metric tensor $g_{\mu}^{\nu}$ and the
dilaton field $\Phi$, the equations of motion for this gravity can be
written as
\begin{eqnarray}
R_{\mu }^{\nu }-\frac{1}{2}g_{\mu }^{\nu }R &=&2\partial _{\mu }\Phi
\partial ^{\nu }\Phi -\frac{1}{2}g_{\mu }^{\nu }V\left( \Phi \right)
\nonumber \\
&&-g_{\mu }^{\nu }\partial _{c}\Phi \partial ^{c}\Phi +KT_{\mu }^{\nu },
\label{Eq1} \\
\nabla ^{2}\Phi &=&\frac{1}{4}\frac{\partial V}{\partial \Phi },  \label{Eq2}
\end{eqnarray}%
where $K=\frac{8\pi G}{c^{4}}$. In order to construct consistent
solutions of the field equations (\ref{Eq1}) and (\ref{Eq2}), we
adopt the approach of gravitational papers \cite{Chan} and assume
that the dilaton potential contains two Liouville terms
\begin{equation}
V(\Phi )=2\Lambda _{0}e^{2\xi _{0}\Phi }-2\Lambda e^{2\xi \Phi },
\end{equation}
where $\Lambda _{0}$, $\Lambda $, $\xi _{0}$, and $\xi $ are constants. This
kind of potential was previously investigated in \cite{Chan}.

In the present work, we want to obtain the static solutions of Eq.
(\ref{Eq1}). So, we assume the spacetime metric has the following
form
\begin{equation}
ds^{2}=f(r)dt^{2}-\frac{dr^{2}}{g(r)}-r^{2}R^{2}(r)\left( d\theta ^{2}+\sin
^{2}\theta d\varphi ^{2}\right) ,  \label{metric}
\end{equation}
where $f(r)$, $g(r)$\ and $R(r)$ are functions of $r$ which should be
determined.

The equations (\ref{Eq1}) and (\ref{Eq2}) contain four unknown functions $%
f(r)$, $g(r)$, $R(r)$ and $\Phi(r)$. In order to solve them, we consider the
ansatz
\begin{equation}
R(r)=e^{\alpha \Phi (r)}.  \label{R(r)}
\end{equation}

This ansatz was first introduced in \cite{DehghaniII} for the purpose of
finding black string solutions of Einstein Maxwell dilaton gravity. It is
notable that in the absence of the dilaton field ($\alpha =0$ and so $R(r)=1$%
), dilaton gravity reduces to EN gravity. Also, using Eqs. (\ref{Eq2}) and (%
\ref{R(r)}) and the metric introduced in Eq. (\ref{metric}), we can obtain
\begin{equation}
\Phi (r)=\frac{\alpha }{\mathcal{K}_{1,1}}\ln \left( \frac{b}{r}\right) ,
\label{Phi(r)}
\end{equation}
where $b$ is an arbitrary constant and $\mathcal{K}_{i,j}=i+j\alpha ^{2}$.

On the other hand, the energy-momentum tensor for a perfect fluid is
\begin{equation}
T^{\mu \nu }=\left( P+\rho c^{2}\right) U^{\mu }U^{\nu }+Pg^{\mu \nu },
\label{EMTensorEN}
\end{equation}
where $P$ and $\rho $ are, respectively, pressure and density of the fluid
which are measured by the local observer, and $U_{\mu }$ is the fluid
four-velocity. Using Eq. (\ref{EMTensorEN}) and the metric introduced in Eq.
(\ref{metric}), we can obtain the components of energy-momentum for $(3+1)$%
-dimensions as follows
\begin{equation}
T_{0}^{0}=\rho c^{2} \hspace{1cm} \& \hspace{1cm}
T_{1}^{1}=T_{2}^{2}=T_{3}^{3}=-P.  \label{4dim}
\end{equation}

Now, we consider the metric (\ref{metric}) and Eq. (\ref{4dim}) for the
perfect fluid to obtain the components of Eq. (\ref{Eq1}) with the following
forms
\begin{eqnarray}
K\rho c^{2} &=&\frac{\left( 1-r^{2}\Lambda \Upsilon ^{2\alpha ^{2}}\right) }{%
\Upsilon ^{\alpha ^{2}}r^{2}}+\frac{\alpha ^{2}\Upsilon }{b^{2}\mathcal{K}%
_{-1,1}}  \nonumber \\
&&-\frac{\left( g+rg{^{\prime }}\mathcal{K}_{1,1}\right) }{r^{2}\mathcal{K}%
_{1,1}^{2}},  \label{I} \\
KP &=&-\frac{\left( 1-r^{2}\Lambda \Upsilon ^{2\alpha ^{2}}\right) }{%
\Upsilon ^{\alpha ^{2}}r^{2}}-\frac{\alpha ^{2}\Upsilon }{b^{2}\mathcal{K}%
_{-1,1}}  \nonumber \\
&&+\frac{g\left( f+rf{^{\prime }}\mathcal{K}_{1,1}\right) }{r^{2}f\mathcal{K}%
_{1,1}^{2}},  \label{II} \\
KP &=&-\frac{\left( 4f\Lambda \Upsilon ^{\alpha ^{2}}-2gf{^{\prime \prime }}%
-ff{^{\prime }}(\frac{g}{f}){^{\prime }}\right) }{4f}  \nonumber \\
&&-\frac{\alpha ^{2}\Upsilon }{b^{2}\mathcal{K}_{-1,1}}+\frac{(fg){^{\prime }%
}}{2rf\mathcal{K}_{1,1}}+\frac{\alpha ^{2}g}{r^{2}\mathcal{K}_{1,1}^{2}},
\label{III}
\end{eqnarray}
where
\begin{equation}
\Upsilon =\left( \frac{b}{r}\right) ^{\frac{2}{\mathcal{K}_{1,1}}},
\end{equation}
and $f$, $g$, $\rho$ and $P$ are functions of $r$. It is notable
that the prime and double prime are, respectively, the first and
second derivatives with respect to $r$. On the other hand,
substituting $\alpha=0$ in Eqs. (\ref{I}-\ref{III}), one finds the
corresponding field equations of EN gravity (see \cite{BordbarHE}
for more details).

Using Eqs. (\ref{I}--\ref{III}) and after some algebraic calculations, we
obtain
\begin{eqnarray}
&&\frac{dP}{dr}+\frac{f{^{\prime }}\left( c^{2}\rho +P\right) \mathcal{K}%
_{1,-1}}{2f}+\frac{2\alpha ^{2}\left( 1-r^{2}\Lambda \Upsilon ^{2\alpha
^{2}}\right) }{K\Upsilon ^{\alpha ^{2}}r^{3}\mathcal{K}_{1,1}}-  \nonumber \\
&&\frac{2\alpha ^{2}\Upsilon }{rKb^{2}\mathcal{K}_{1,1}\mathcal{K}_{-1,1}}+%
\frac{\alpha ^{2}gf{^{\prime \prime }}}{rf\mathcal{K}_{1,1}}+\frac{\alpha
^{2}\left( rg{^{\prime }}+2g\right) }{r^{3}\mathcal{K}_{1,1}^{2}}=0.
\label{extraEQ}
\end{eqnarray}

In addition, one can use Eq. (\ref{II}) to obtain $f{^{\prime}}$ with the
following form
\begin{eqnarray}
f{^{\prime }} &=&\frac{rf\left( \frac{\Upsilon ^{-\alpha ^{2}}}{r^{2}g}%
+\Lambda \Upsilon ^{\alpha ^{2}}+KP\right) \mathcal{K}_{1,1}}{g}  \nonumber
\\
&&+\frac{r\alpha ^{2}\Upsilon \mathcal{K}_{1,1}f}{gb^{2}\mathcal{K}_{-1,1}}-%
\frac{f}{r\mathcal{K}_{1,1}}.  \label{diff(r)}
\end{eqnarray}

To obtain the function $g(r)$, we consider Eq. (\ref{I}). After integration
we achieve
\begin{eqnarray}
g(r) &=&\left( \frac{\alpha ^{2}\Upsilon }{b^{2}\mathcal{K}_{1,2}\mathcal{K}%
_{1,-1}}+\frac{\Lambda \Upsilon ^{\alpha ^{2}}}{3}+\frac{1}{\Upsilon
^{\alpha ^{2}}r^{2}\mathcal{K}_{1,2}}\right) \mathcal{K}_{1,1}^{2}r^{2}
\nonumber \\
&&-\frac{Kc^{2}\mathcal{K}_{1,1}}{r^{\frac{1}{\mathcal{K}_{1,1}}}}\int \rho
(r,\alpha )r^{\frac{\mathcal{K}_{2,1}}{\mathcal{K}_{1,1}}}dr,
\label{ggtotall}
\end{eqnarray}%
where $\rho (r,\alpha )=\frac{dM\left( r,\alpha \right) }{dV_{eff}}$, in
which $V_{eff}=\frac{4}{3}\pi R_{eff}^{3}$ and $R_{eff}=\left( \frac{3%
\mathcal{K}_{1,1}}{\mathcal{K}_{2,3}}\right) ^{1/3}r^{\mathcal{K}_{2,3}/3%
\mathcal{K}_{1,1}}$. It is interesting to note that in the
presence of dilaton field we should replace $r$ with $R_{eff}$. In
other words, dilaton field can modify the radius of sphere into
$R_{eff}$ instead of usual sphere with radius $r$. It is notable
that, when $\alpha =0$ (in the absence of the dilaton field),
$R_{eff}$ reduces to $r$ and also (as we expect) we obtain
$g(r)=1-$ $\frac{\Lambda }{3}r^{2}-\frac{m(r)}{4\pi r}$, where
$m(r)=\int 4\pi r^{2}\rho (r)dr$.

Now, we consider the integral that appears in Eq. (\ref{ggtotall})
and by using of $R_{eff}$, the equation (\ref{ggtotall}) turns
into
\begin{eqnarray}
g(r) &=&\left( \frac{\alpha ^{2}\Upsilon }{b^{2}\mathcal{K}_{1,2}\mathcal{K}
_{1,-1}}+\frac{\Lambda \Upsilon ^{\alpha ^{2}}}{3}+\frac{1}{\Upsilon
^{\alpha ^{2}}r^{2}\mathcal{K}_{1,2}}\right) \mathcal{K}_{1,1}^{2}r^{2}
\nonumber \\
\nonumber \\
&&-\frac{Kc^{2}\mathcal{K}_{1,1}}{4\pi r^{\frac{1}{\mathcal{K}_{1,1}}}}
M_{eff}\left( r,\alpha \right) ,  \label{gggtotall}
\end{eqnarray}%
where we used $M_{eff}\left( r,\alpha \right) =\int \rho (r,\alpha
)4\pi R_{eff}^{2}dR_{eff}$. It is notable that, $M_{eff}\left(
r,\alpha \right) $ and $R_{eff}$ are, respectively, the effective
mass and radius as results of the presence of the dilaton field.
In obtained solution (\ref{gggtotall} ) and for consistency we use
\begin{equation}
\xi _{0}=\frac{1}{\alpha },\ \ \ \ \ \ \xi =\alpha ,\ \ \ \ \ \ \Lambda _{0}=%
\frac{\alpha ^{2}}{b^{2}\mathcal{K}_{-1,1}}.  \nonumber
\end{equation}

Notice that $\Lambda $ remains as a free parameter which plays the role of
the cosmological constant.

Now, we can obtain the HEE for dilaton gravity. For this purpose, we
consider the Eqs. (\ref{diff(r)}) and (\ref{gggtotall}), and inserting them
in Eq. (\ref{extraEQ}). After some manipulation we obtain
\begin{equation}
\frac{dP}{dr}=\frac{\alpha ^{2}\left( \mathcal{A}c^{2}-\mathcal{B} \mathcal{K%
}_{1,1}^{2} +\mathcal{C}+\mathcal{D}\right) }{Kr^{3}f\mathcal{K}_{1/2,1}%
\mathcal{K}_{-1,1}\mathcal{K}_{1,1}^{3}},  \label{TOV}
\end{equation}%
where $\mathcal{A}$, $\mathcal{B}$, $\mathcal{C}$ and $\mathcal{D}$ are in
the following forms
\begin{eqnarray}
\mathcal{A} &=&\frac{K}{4\pi r^{1/\mathcal{K}_{1,1}} }\mathcal{K}_{1,1}%
\mathcal{K}_{-1,1}\mathcal{K}_{1/2,1}\mathcal{X}_{1,2,1}M_{eff}\left(
r,\alpha \right) , \nonumber\\
\mathcal{B} &=&\frac{r^{2}}{3}\left( \Lambda \mathcal{K}_{-1,1}\mathcal{K}%
_{1/2,1}\gamma ^{\alpha ^{2}}\mathcal{X}_{2,1,2}+\frac{3\alpha
^{2}\gamma \mathcal{X}_{1,1,2} }{2b^{2}}\right) , \nonumber \\
\mathcal{C} &=&2\gamma ^{\alpha ^{2}}\mathcal{K}_{1,1}^{2}\mathcal{K}%
_{-1,1}\left( \Lambda r^{2}f\mathcal{K}_{1/2,1}-\frac{\mathcal{X}_{1,2,4}}{%
4\gamma ^{2\alpha ^{2}}}\right) , \nonumber \\
\mathcal{D}
&=&r^{2}f\mathcal{K}_{1,1}^{2}\mathcal{K}_{1/2,1}\left[
Kc^{2}\rho \mathcal{K}_{-1,1}^{2}+\left( P+\rho c^{2}\right) \mathcal{Y}%
\right] ,\nonumber
\end{eqnarray}%
and also $\mathcal{X}_{i,j,k}=r^{2}\mathcal{K}_{1,1}^{2}f{^{\prime \prime }+k%
}\mathcal{K}_{i,j}f$ \ and $\mathcal{Y=}\frac{2\gamma }{b^{2}}+\frac{Krf{%
^{\prime }}}{2\alpha ^{2}f}\mathcal{K}_{1,1}^{2}\mathcal{K}_{-1,1}^{2}$.

In next section, we continue our paper with considering dilaton gravity as a
correction (perturbation) of Einstein gravity and we will obtain the
corresponding HEE.

\section{Dilaton gravity as a correction of Einstein gravity}

The interesting agreements and acceptable results of the EN
gravity with experimental results (observations) guide us to
consider its modification (such as dilaton gravity) as a
correction of EN gravity. On the other hand, to avoid the
complexity of modified gravity theories and obtaining credible
solutions, it is logical to consider the dominant perturbative
terms. Therefore in this section, we discuss the effects of small
value $\alpha $. When $\alpha $ is very small, we can use series
expansion in Eq. (\ref{TOV}) and keep the dominant contribution
term. So, we keep $\mathcal{O}(\alpha ^{2})$ and ignore
$\mathcal{O}(\alpha ^{4})$ and higher orders to obtain
\begin{eqnarray}
\frac{dP}{dr} &=&\frac{\left[ 3Kc^{2}m(r)+4\pi r^{3}\left( 3KP+2\Lambda
\right) \right] }{r^{2}\left[ 3Kc^{2}m(r)-4\pi r\left( \Lambda
r^{2}+3\right) \right] }\left( P+\rho c^{2}\right) +  \nonumber \\
&&\frac{3\left( P+\rho c^{2}\right) \mathcal{H}\alpha ^{2}}{\mathcal{K}_{1,1}%
\left[ r\left( \Lambda r^{2}+3\right) -\frac{3Kc^{2}m(r)}{4\pi }\right] ^{2}}%
+\mathcal{O}(\alpha ^{4}),  \label{TOVcorrection}
\end{eqnarray}%
where $\mathcal{H}$ is
\begin{eqnarray}
\mathcal{H} &=&\frac{3Kc^{2}\left[ 1+r^{2}\left( \Lambda +KP\right) \right] %
\left[ 4\pi \mathcal{M}(r)-m(r)\right] }{8\pi }  \nonumber \\
&&+\frac{9K^{2}c^{6}m^{3}(r)}{64\pi ^{3}r^{4}\left( P+\rho c^{2}\right) }-%
\frac{3Kc^{2}m(r)\varpi _{1}}{4\pi }  \nonumber \\
&&-\frac{3Kc^{4}m^{2}(r)\left[ 12+r^{2}\left( 2\Lambda +KP\right) \right] }{%
16\pi ^{2}r^{3}\left( P+\rho c^{2}\right) }+\varpi _{2},  \nonumber
\end{eqnarray}%
in which%
\begin{eqnarray*}
\varpi _{1} &=&\frac{P\left[ \ln r-5-\Lambda r^{2}\left( \frac{10}{3}-\ln
r\right) \right] }{2\left( P+\rho c^{2}\right) } \\
&&+\frac{c^{2}\rho \left[ \left( \Lambda r^{2}+1\right) \ln r-2\Lambda
r^{2}-1\right] }{2\left( P+\rho c^{2}\right) } \\
&&-\left( r^{2}\Lambda -1\right) \ln \left( \frac{b}{r}\right)
+Kr^{2}P\left( \frac{\ln r}{2}-1\right)  \\
&&-\frac{18\Lambda r^{2}+\Lambda ^{2}r^{4}+45}{3Kr^{2}\left( P+\rho
c^{2}\right) }, \\
&& \\
\varpi _{2} &=&r^{3}\left( c^{2}\rho +P\right) \left[ KP\left( \Lambda
r^{2}-3\right) -4\Lambda \right] \ln \left( \frac{b}{r}\right)  \\
&&-\frac{Pr\left[ \frac{2\Lambda ^{2}r^{4}}{3}+9Kc^{2}r^{2}\rho +12\left(
1+\Lambda r^{2}\right) \right] }{2} \\
&&-\frac{2\left( \Lambda r^{2}+3\right) ^{2}}{Kr}-\frac{9Kr^{3}P^{2}}{2} \\
&&-c^{2}r\rho \left( 4\Lambda r^{2}+3\right) , \\
&& \\
\mathcal{M}(r) &=&\int r^{2}\ln (r)\rho (r)dr.
\end{eqnarray*}%
It is notable that, as one expects the first term of Eq. (\ref%
{TOVcorrection}) is the TOV equation in the presence of
cosmological constant. In addition, the second term
($\mathcal{H}$) is the leading order term of considered dilaton
field as a correction to EN gravity.

We are going to continue our paper with considering HEE obtained in Eq. (\ref%
{TOVcorrection}) and obtain the properties of neuron stars by using of
cluster expansion of the energy.

\section{Structure properties of neutron star}

\subsection{Equation of state of neutron star matter}

In order to study properties of the neutron star structure, we
need to consider an equation of state for neutron star matter. The
constituents of interior part of a neutron star are neutrons,
protons, electrons and muons which are in charge
neutrality and beta equilibrium conditions (beta-stable matter) \cite%
{Shapiro}. In recent years, we have obtained the equation of state of
neutron star matter using the microscopic constrained variational
calculations based on the cluster expansion of the energy functional \cite%
{BordbarRi,Bordbar,BordbarH}. In these studies the modern
two-nucleon potentials such as the new Argonne $AV_{18}$
\cite{Wiringa} and charged dependent Reid-$93$\ \cite{Stoks} have
been used. One of the importances of these calculations is the
fact that it does not bring any free parameter into the formalism,
and its results show a good convergence. In this method, a
microscopic computation of asymmetry energy is carried on for the
asymmetric nuclear matter calculations which results into more
accuracy in comparison with other semi-empirical parabolic
approximation methods. In fact, a microscopic calculation is
required with the modern nucleon-nucleon potentials which are
explicitly depend on the isospin projection ($T_{z}$)
\cite{BordbarM}.

\begin{figure}[tb]
\includegraphics[width=7cm]{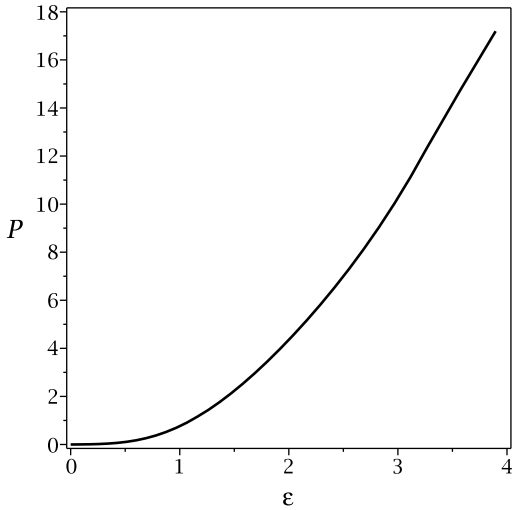}
\caption{ Equation of state of neutron star matter (pressure, $P$
($10^{35}$ erg/$cm^3$) versus density, $\epsilon$ ($10^{15}$
gr/$cm^3$)).} \label{Fig1}
\end{figure}

In this paper, for neutron star structure, we use the modern
equation of state which has been calculated using $AV_{18}$
potential \cite{BordbarRi,Bordbar,BordbarH} and investigate some
physical properties of neutron star structure. This equation of
state of neutron star matter is shown in Fig. \ref{Fig1}.

%%%%%%%%%%%%%%%%%%%%%%%%%%%%%%%%%%%%%%%%%%%%%%%%%%%%%%%%%%%%%%%%

%%%%%%%%%%%%%%%%%%%%%%%%%%%%%%%%%%%%%%%%%%%%%%%%%%%%%%%%%%%%%%%%%%%

\subsection{Maximum mass of neutron star}

In order to distinguish neutron stars from black holes, it is
important to know the maximum gravitational mass of a neutron star
for stability against collapse into a black hole. In other words,
it is expected that below a certain maximum mass, degenerate
pressure due to the nucleons is sufficient to prevent an object
from becoming a black hole \cite{Shapiro}. Therefore, determining
the maximum gravitational mass of neutron stars is of special
importance in astrophysics. Direct ways to measure the neutron
star mass are observations of the X-ray pulsars and X-ray
bursters, but there are many errors in these methods and so,
measuring the mass of neutron star is not very useful and
accurate. Fortunately, the mass of neutron stars have been
determined with high accuracy using the binary radio pulsars \cite
{Weisberg,Liang,Heap,Jonker,Quaintrell}.

Here, we calculate the maximum mass of neutron star using the
equation of state of neutron star matter presented in Fig.
\ref{Fig1} by numerical integrating the HEE obtained in Eq.
(\ref{TOVcorrection}). This leads to the neutron star mass and
radius as a function of central mass density ($\epsilon _{c}$).
For this purpose, by selecting a central mass density
($\epsilon_{c}$), under the boundary conditions $P(r=0)=P_{c}$ and
$m(r=0)=0$, we integrate the Eq. (\ref{TOVcorrection}) outwards to
a radius $r=R$ in which $P$ vanishes. This yields the neutron star
radius $R$ and mass $M=m(R)$. Our results have been given in the
following figures and tables.

%%%%%%%%%%%%%%%%%%%%%%%%%%%%%%%%%%%%%%%%%%%%%%%%%%%%%%%%%%%%%%%%%%%%%%%%%%%%%%%%%%%%%%%%%%%%%%%%%%%%%%%%%%%%%%%%%%%%%%%%%%%%%%
\begin{table}
\begin{center}
\caption{Maximum mass of neutron star and its corresponding radius
for various values of
 $\alpha$ at $\Lambda =0$ and $b=10^{-2}$.}
\begin{tabular}{ccc}
\hline\hline $\alpha $ & ${M_{max}}\ (M_{\odot})$ & $R\ (km)$ \\
\hline\hline $1.33\times 10^{-11}$ & $0.22$ & $12.60$ \\ \hline
$1.30\times 10^{-11}$ & $0.81$ & $10.01$ \\ \hline $1.20\times 10^{-11}$ & $1.37$ & $9.62$ \\
\hline $1.10\times 10^{-11}$ & $1.53$ & $9.33$ \\ \hline
$1.00\times 10^{-11}$ & $1.60$ & $9.11$  \\ \hline $5.00\times
10^{-12}$ & $1.68$ & $8.55$
\\ \hline $1.00\times 10^{-12}$ & $1.68$ & $8.42$ \\ \hline \label{tab1}
\end{tabular}
\end{center}
\end{table}
%%%%%%%%%%%%%%%%%%%%%%%%%%%%%%%%%%%%%%%%%%%%%%%%%%%%%%%%%%
%%%%%%%%%%%%%%%%%%%%%%%%%%%%%%%%%%%%%%%%%%%%%%%%%%%%%%%%%%%%%%%%%%

\begin{table}
\begin{center}
\caption{Maximum mass of neutron star and its corresponding radius
for various values of $\alpha$ and $\Lambda$ at $b=10^{-2}$.}
\begin{tabular}{cccc}
\hline\hline $\alpha $ & ${M_{max}}\ (M_{\odot})$ & $R\ (km)$ &
$\Lambda $ \\ \hline\hline $1.00\times 10^{-12}$ & $0.77$ & $6.64$
& $1.00\times 10^{-11}$ \\ \hline $1.00\times 10^{-13}$ & $0.78$ &
$6.65$ & $1.00\times 10^{-11}$ \\ \hline $1.00\times 10^{-14}$ &
$0.78$ & $6.65$ & $1.00\times 10^{-11}$ \\ \hline $1.00\times
10^{-12}$ & $1.55$ & $8.25$ & $1.00\times 10^{-12}$ \\ \hline
$1.00\times 10^{-13}$ & $1.56$ & $8.25$ & $1.00\times 10^{-12}$ \\
\hline $1.00\times 10^{-11}$ & $1.56$ & $9.04$ & $1.00\times 10^{-13}$ \\
\hline $1.00\times 10^{-13}$ & $1.67$ & $8.40$ & $1.00\times 10^{-13}$ \\
\hline $1.00\times 10^{-14}$ & $1.67$ & $8.40$ & $1.00\times 10^{-13}$ \\
\hline $1.00\times 10^{-11}$ & $1.59$ & $9.10$ & $1.00\times 10^{-14}$ \\
\hline $1.00\times 10^{-12}$ & $1.68$ & $8.42$ & $1.00\times 10^{-14}$ \\
\hline $1.00\times 10^{-13}$ & $1.68$ & $8.42$ & $1.00\times 10^{-14}$ \\
\hline \label{tab2}
\end{tabular}
\end{center}
\end{table}

%%%%%%%%%%%%%%%%%%%%%%%%%%%%%%%%%%%%%%%%%%%%%%%%%%%%%%%%%%%%%%%%%%

%%%%%%%%%%%%%%%%%%%%%%%%%%%%%%%%%%%%%%%%%%%%%%%%%%%%%%%%%%%%%%%%%%%%%%%%%%%%%%%%%%%%%%%%%%%%%%%%%%%%%%%%%%%%%%%%%%%%%%%%%%%%%%
%%%%%%%%%%%%%%%%%%%%%%%%%%%%%%%%%%%%%%%%%%%%%%%%%%%%%%%%%%%%%%%
\begin{figure}[tb]
$
\begin{array}{c}
\includegraphics[width=7cm]{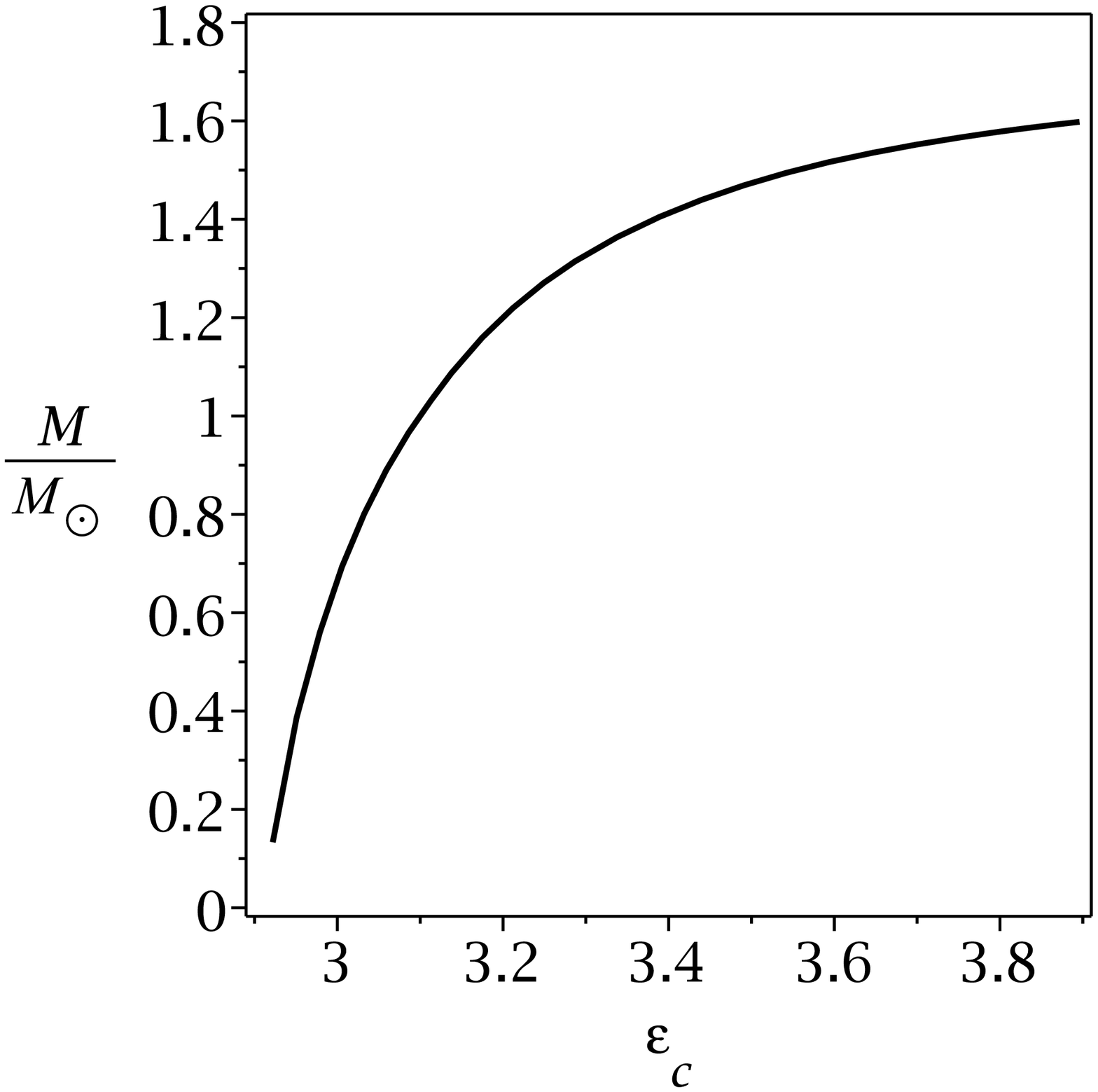}\\
\includegraphics[width=7cm]{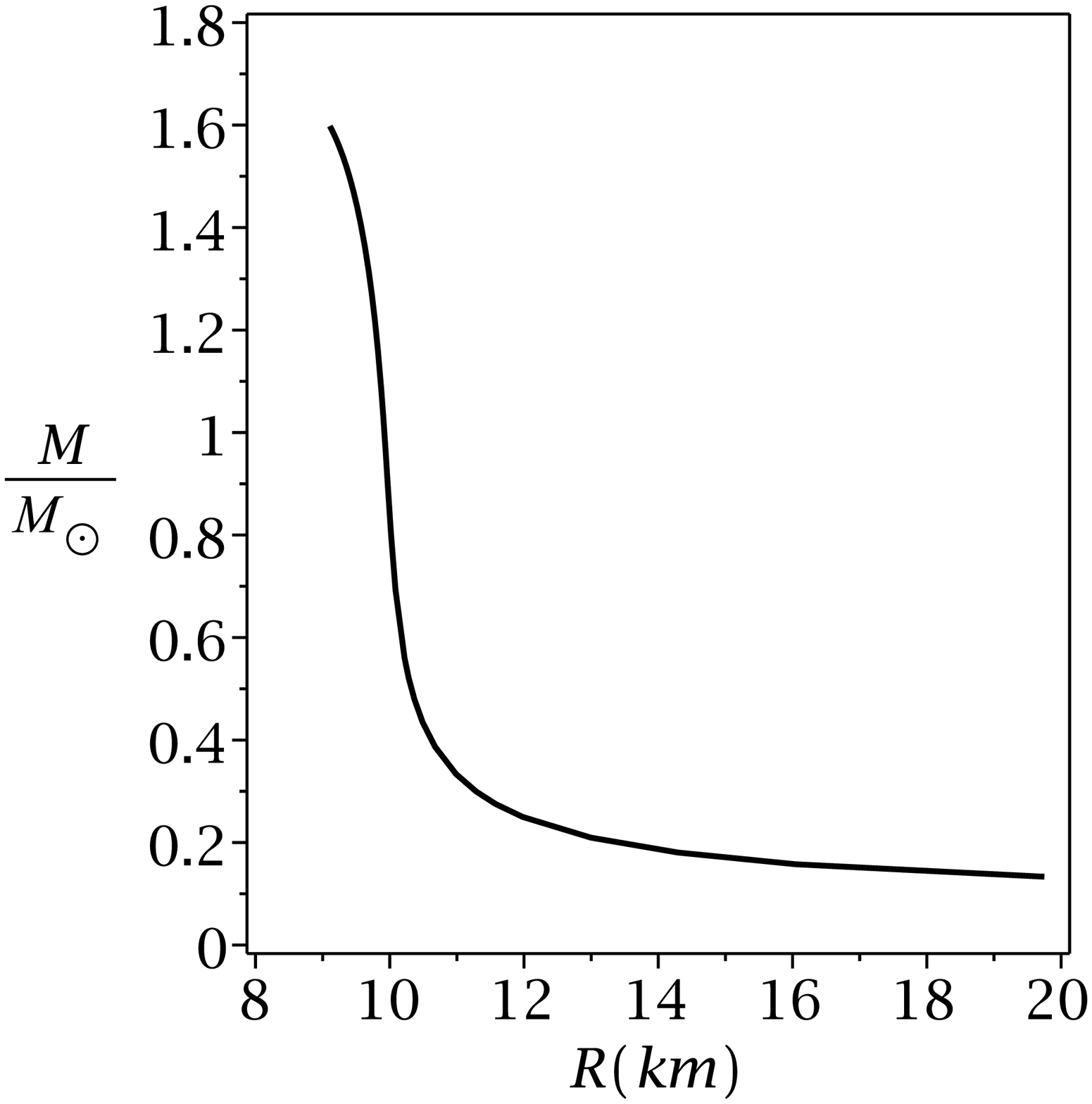}
\end{array}
$%
\caption{Gravitational mass versus central mass density,
$\epsilon_{c}$ ($10^{15}$gr/$cm^3$), (up) and radius (down) for
$b=10^{-2}$,
$\Lambda =0$ and $\protect\alpha %
=10^{-12} $.} \label{Fig2}
\end{figure}
%%%%%%%%%%%%%%%%%%%%%%%%%%%%%%%%%%%%%%%%%%%%%%%%%%%%%%%%%%%%%%%
%%%%%%%%%%%%%%%%%%%%%%%%%%%%%%%%%%%%%%%%%%%%%%%%%%%%%%%%%%%%%%%
\begin{figure}[tb]
$%
\begin{array}{cc}
\includegraphics[width=7cm]{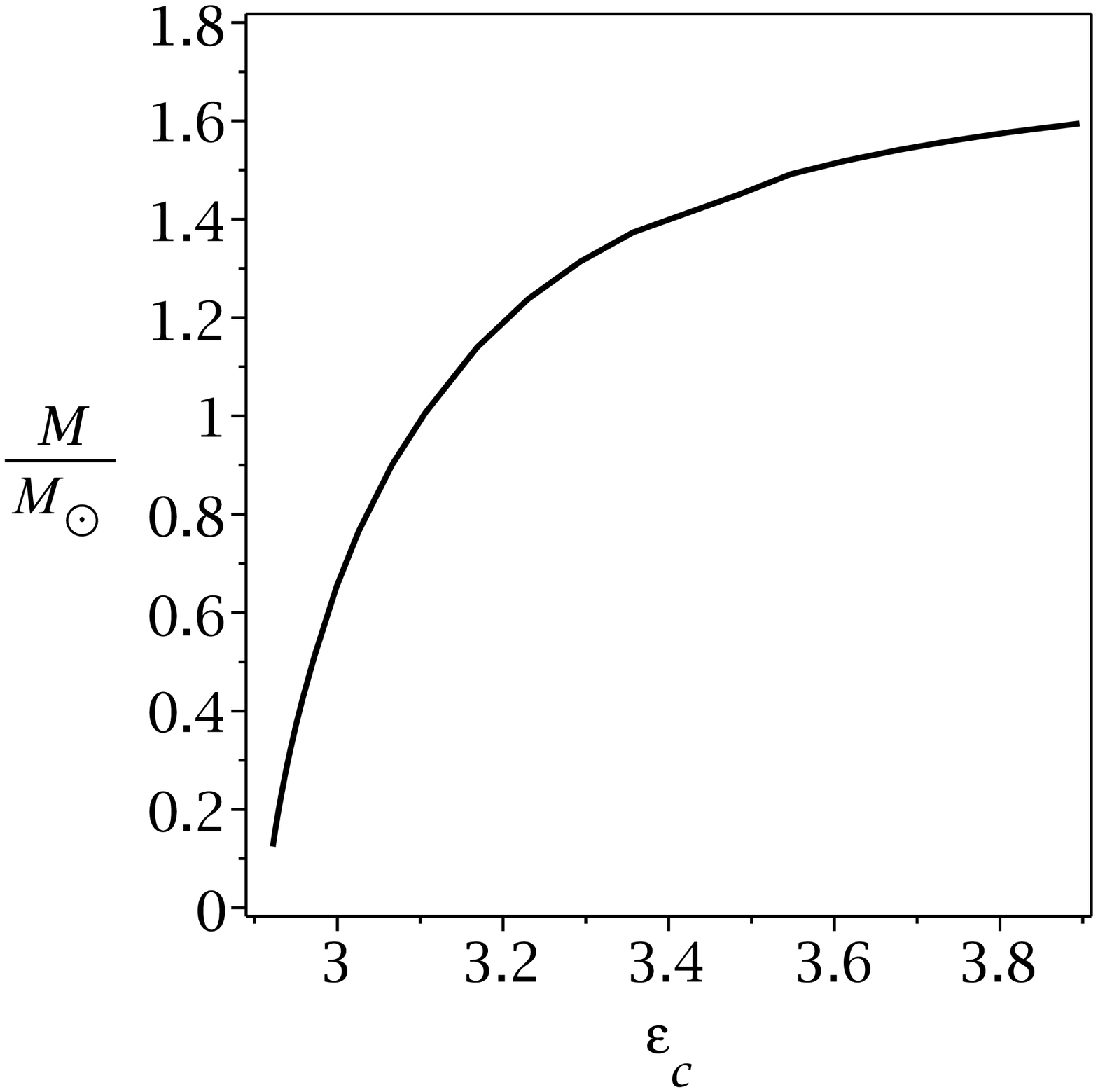}\\
\includegraphics[width=7cm]{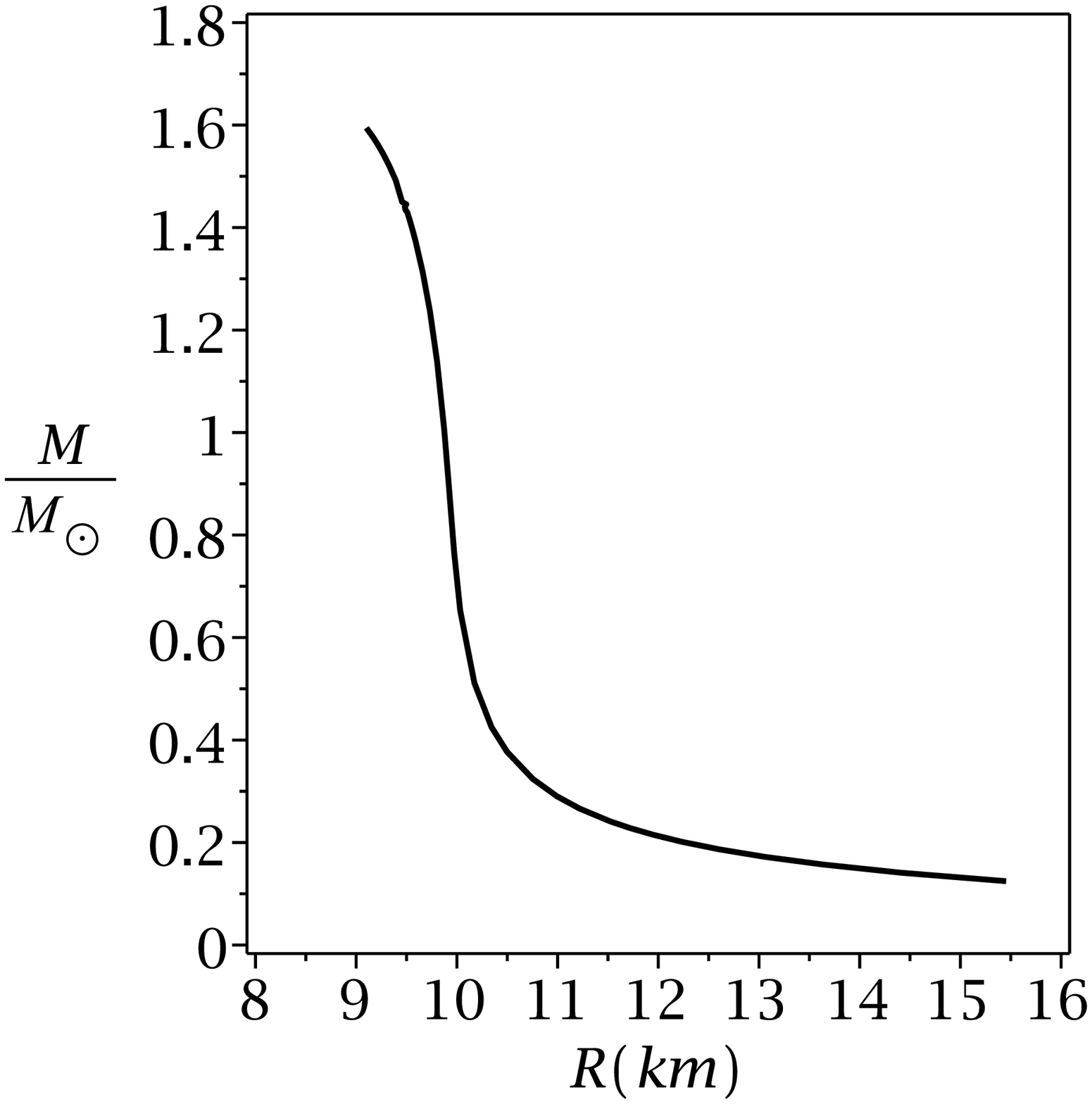}
\end{array}
$%
\caption{Gravitational mass versus central mass density,
$\epsilon_{c}$ ($10^{15}$gr/$cm^3$), (up) and radius (down) for
$b=10^{-2}$, $\Lambda =10^{-14}$ and $\protect\alpha=10^{-11}$.}
\label{Fig3}
\end{figure}
%%%%%%%%%%%%%%%%%%%%%%%%%%%%%%%%%%%%%%%%%%%%%%%%%%%%%%%%%%%%%%
%%%%%%%%%%%%%%%%%%%%%%%%%%%%%%%%%%%%%%%%%%%%%%%%%%%%%%%%%%%%%%%
\begin{figure}[tb]
$%
\begin{array}{cc}
\includegraphics[width=7cm]{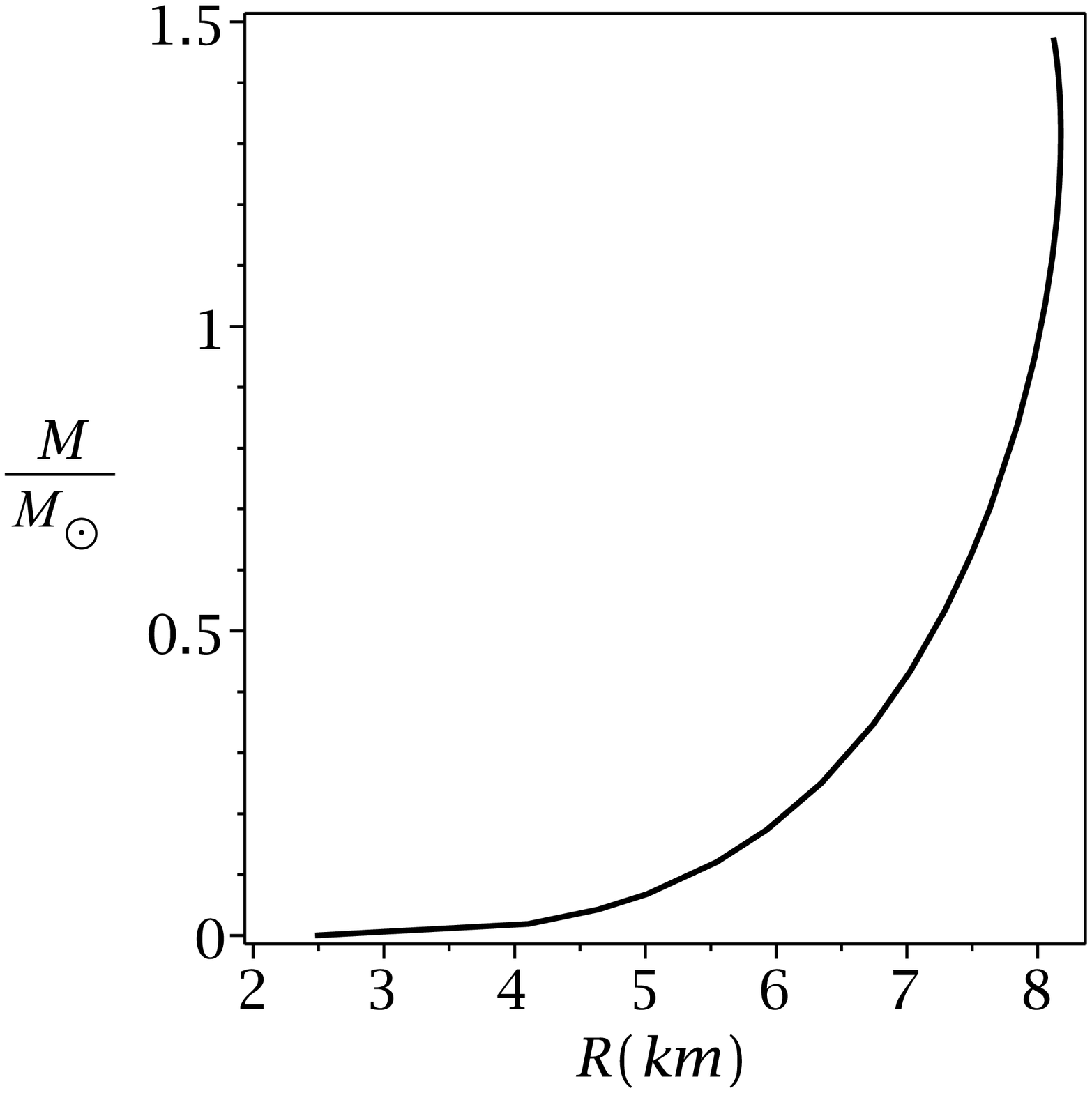}\\
\includegraphics[width=7cm]{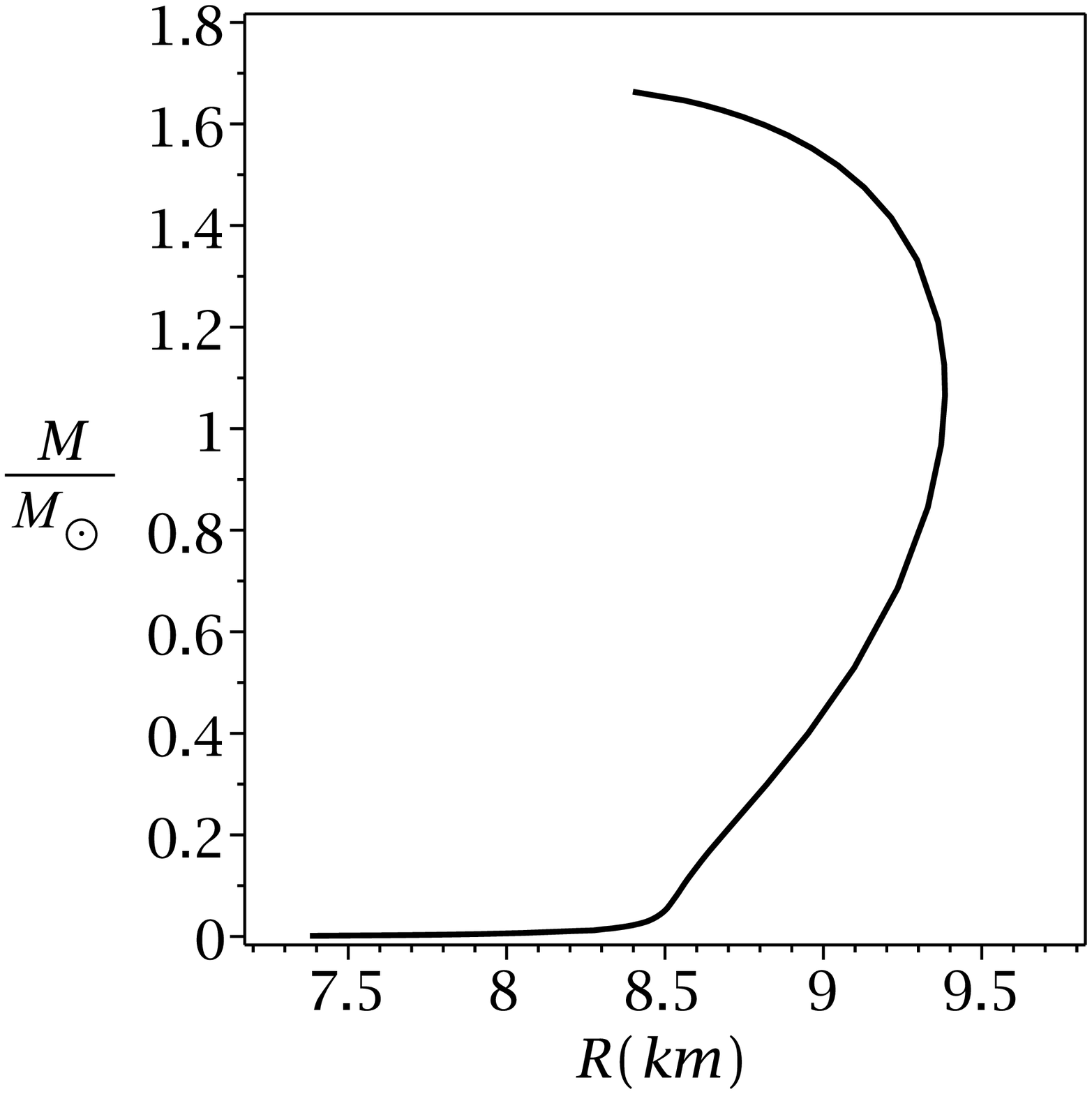}
\end{array}
$%
\caption{ Gravitational mass versus radius for $b=10^{-2}$ and $%
\protect\alpha =10^{-12}$ at $\Lambda =1.7 \times 10^{-12}$ (up)
and $\Lambda =1.7 \times 10^{-13}$ (down).} \label{Fig4}
\end{figure}
%%%%%%%%%%%%%%%%%%%%%%%%%%%%%%%%%%%%%%%%%%%%%%%%%%%%%%%%%%%%%%%
%%%%%%%%%%%%%%%%%%%%%%%%%%%%%%%%%%%%%%%%%%%%%%%%%%%%%%%%%%%%%%%%%%
\begin{table}
\begin{center}
\caption{Maximum mass of neutron star and its corresponding radius
for various values of $\alpha$ and (negative) $\Lambda$ at
$b=10^{-2}$.}
\begin{tabular}{cccc}
\hline\hline $\alpha $ & ${M_{max}}\ (M_{\odot})$ & $R\ (km)$ &
$\Lambda $ \\ \hline\hline $1.00\times 10^{-13}$ & $1.77$ & $8.62$
& $-1.00\times 10^{-12}$ \\ \hline $1.00\times 10^{-12}$ & $1.78$
& $8.63$ & $-1.00\times 10^{-12}$ \\ \hline $1.00\times 10^{-11}$
& $2.25$ & $10.57$ & $-1.00\times 10^{-12}$ \\ \hline $1.00\times
10^{-13}$ & $1.70$ & $8.44$ & $-1.00\times 10^{-13}$ \\ \hline
$1.00\times 10^{-12}$ & $1.69$ & $8.43$ & $-1.00\times 10^{-13}$ \\
\hline $1.00\times 10^{-11}$ & $1.63$ & $9.17$ & $-1.00\times 10^{-13}$ \\
\hline \label{tab3}
\end{tabular}
\end{center}
\end{table}

%%%%%%%%%%%%%%%%%%%%%%%%%%%%%%%%%%%%%%%%%%%%%%%%%%%%%%%%%%%%%%%%%%%%%%%%%%%%%%%%%%%%%%%%%%%%%%%%%%%%%%%%%%%%%%%%%%%%%%%%%%%%%%

Our results show that the maximum mass and radius of neutron stars
depend on the parameters of dilaton field ($\alpha$) and
cosmological constant ($\Lambda$). As one can see in Table
\ref{tab1}, we have obtained the maximum mass and radius of
neutron star in the absence of cosmological constant. The results
show that as $\alpha$ increases, the maximum mass of neutron star
decreases. We have found that the maximum mass of neutron star is
affected with variation of $\alpha$ for the case of
$\alpha>10^{-12}$. In other words, we can omit the effects of
$\alpha$ for values lower than about $10^{-12}$, where these
results are consistent with those were obtained in Ref.
\cite{BordbarH}.

Considering $\alpha=10^{-12}$ and $\Lambda =0$, we plot the
neutron star gravitational mass (in solar mass unit $M_{\odot}$)
as a function of central mass density ($\epsilon _{c}$) in Fig.
\ref{Fig2} (up diagram). Our results show that at low densities,
the calculated neutron star mass exhibits a minimum ($\approx 0.12
M_{\odot}$). It can be seen that at high densities, the increasing
of gravitational mass becomes very slow, and finally it approaches
a limiting value ($\approx 1.68 M_{\odot}$). This limiting value
is the maximum gravitational mass of neutron star, and its
corresponding central density is the highest possible value for
the neutron star central density. A star with higher central
density would be unstable against the gravitational collapse to a
black hole. For mentioned quantities of $\alpha$ and $\Lambda$
($\alpha =10^{-12}$ and $\Lambda =0$), we also plot the
gravitational mass versus radius for in Fig. \ref{Fig2} (down
diagram). Our results show that for the neutron star, there are a
minimum gravitational mass ($\approx  0.12 M_{\odot}$) and a
maximum gravitational mass ( $\approx 1.68 M_{\odot}$). It is well
known that in this mass region, the equilibrium configuration of
neutron stars can exist.

On the other hand, when we consider the effects of both $\alpha$
and $\Lambda$ in structure of neutron stars, the results are
interesting. Our calculations show that considering a fixed value
for $\Lambda$, the maximum mass of the neutron star increases as
$\alpha$ decreases. Also, when $\alpha$ is a fixed value, by
decreasing $\Lambda$, the maximum mass increases (see Table
\ref{tab2} for more details). Considering the effects of both
$\alpha$ and $\Lambda$, simultaneously, we plot the gravitational
mass versus radius in Fig. \ref{Fig3}. It is notable that for the
special values of $\Lambda$ and $\alpha$, our results show an
abnormality in the behavior of mass versus radius for neutron
star. In other words, for the special values of $\Lambda$ and
$\alpha$, we encounter with an interesting behavior. For these
cases, with increasing radius, the gravitational mass increases
(see Fig. \ref{Fig4}), while for a neutron star, with increasing
radius, the mass decreases, (see Ref. \cite{AstashenokCOII} for
more details). In other words, there are critical values for
$\Lambda$ and $\alpha$, so that, the diagram related to
gravitational mass versus radius is not similar to the diagram
related to ordinary neutron star.

In order to complete our discussion and motivated by AdS
spacetimes, we consider negative cosmological constant and collect
the results in table \ref{tab3}. We conclude that for a fixed
value of dilaton field, increasing the absolute value of negative
cosmological constant leads to increasing the maximum mass of the
neutron star. In other words, for the negative cosmological
constant, one finds that the fraction of maximum mass per solar
mass can be larger than $2$.

\section{Closing Remarks}

In this paper, we considered a $4$-dimensional spherical symmetric
line element and extracted the hydrostatic equilibrium equation of
stars in dilaton gravity. We found that for $\alpha=0$ limit, the
HEE for dilaton gravity reduces to Einstein-$\Lambda$ one, as one
expects. Then, we regarded dilaton field as a correction of EN
gravity and showed that $\frac{dP}{dr}$ contains usual TOV
equation and an extra dilatonic term.

Considering the HEE obtained in this paper and using our neutron
star matter equation of state, we obtain the structure properties
of neutron star. The results showed that as the parameter of
dilaton gravity ($\alpha$) increases, the maximum mass of neutron
star decreases. In other words, as $\alpha $ decreases, the
effects of dilaton gravity decreases, and the results will be
close to the results obtained in Einstein-$\Lambda$ gravity. These
results indicated that the dilaton field behaves as an external
pressure and prevents the increasing mass of neutron star.

In addition, we regarded negative cosmological constant and found
that for a fixed value of dilaton strength, decreasing $\Lambda$
(increasing its absolute value) leads to increasing the maximum
mass of the neutron star. In other words, for both positive and
negative values of cosmological constant, increasing $\Lambda$
leads to decreasing the gravitational mass. Therefore, one may
regard $\Lambda$ as an external pressure which prevents the
increasing mass of neutron star. From another point of view, it
may be interesting to regard $\Lambda$ as a dynamical pressure
\cite{KubiznakM,HendiPE} and relate it to the results of the
present work.

Finally, it was proved that dilaton gravity can be transformed to
Brans-Dicke theory with a suitable conformal transformation. In
addition, it is well-known that a special class of Brans-Dicke
theory ($w_{BD}= 0$) is mathematically equivalent to the $f(R)$
gravity. So, it is interesting to find a relation between obtained
results of the present paper with the corresponding $f(R)$ gravity
and Brans-Dicke theory. We left these problems for future works.

%%%%%%%%%%%%%%%%%%%%%%%%%%%%%%%%%%%%%%%%%%%%%%%%%%%%%%%%%%%%%%%%%%%%%%%%%%%%%%%%%%%%%%%%%%%%%%%%%%%%%%%

\acknowledgments We would like to thank the anonymous referees for
valuable suggestions. B. E. acknowledges S. Panahiyan and R.
Haghbakhsh for helpful discussions. The authors wish to thank
Shiraz University Research Council. This work has been supported
financially by Research Institute for Astronomy and Astrophysics
of Maragha.

\end{document}